\documentstyle{article}
\begin{document}
\title{Dirty black holes: \\Entropy versus area}
\author{Matt Visser\cite{e-mail}\\
        Physics Department\\
	Washington University\\
	St. Louis\\
	Missouri 63130-4899}
\date{3 March 1993}
\maketitle
\begin{abstract}
Considerable interest has recently been expressed in the entropy
versus area relationship for ``dirty'' black holes --- black holes
in interaction with various classical matter fields, distorted by
higher derivative gravity, or infested with various forms of quantum
hair.  In many cases it is found that the entropy is simply related
to the area of the event horizon: $S =  k A_H/(4\ell_P^2)$.  For
example, the ``entropy = (1/4) area'' law {\em holds} for
Schwarzschild, Reissner-Nordstr\"om, Kerr-Newman, and dilatonic
black holes. On the other hand, the ``entropy = (1/4) area'' law
{\em fails} for: various types of  $(Riemann)^n$ gravity, Lovelock
gravity, and various versions of quantum hair. The pattern underlying
these results is less than clear. This paper systematizes these
results by deriving a general formula for the entropy:
\[
S = {k A_H\over4\ell_P^2}
  + {1\over T_H} \int_\Sigma \;
        \left\{\varrho_L-{\cal L}_E\right\} K^\mu d\Sigma_\mu
  + \int_\Sigma s \; V^\mu d\Sigma_\mu.
\]
($K^\mu$ is the timelike Killing vector, $V^\mu$ the four-velocity
of a corotating observer.) If no hair is present the validity of
the ``entropy = (1/4) area'' law reduces to the question of whether
or not the Lorentzian energy density for the system under consideration
is formally equal to the Euclideanized Lagrangian. \newline
****** To appear in Physical Review D 15 July 1993 ******
\end{abstract}

\newpage
\section{INTRODUCTION}

For a variety of reasons, considerable interest has recently been
expressed in the entropy versus area relationship for generic
``dirty''  black holes. By a dirty black hole I mean a black hole
possibly in interaction with various classical matter fields,
possibly modified by higher curvature terms in the gravity Lagrangian
[$(Riemann)^n$], or possibly infested with some version of quantum
hair. Some of these reasons are the following. (1) The low-energy
point-field limit of string theory includes a dilaton field. The
presence of the dilaton field modifies the Reissner-Nordstr\"om and
Kerr-Newman black holes. (2) Despite the successes of string
theory, a fully satisfactory theory of quantum gravity has proved
elusive.  Nevertheless, whatever the underlying quantum theory is,
one would expect on general grounds that the low energy theory
should be describable by the Einstein-Hilbert action modified by
higher-order terms in the Riemann tensor. (3) Quantum hair is a
result of quantum fluctuations in the various low-energy quantum
fields with which the black hole geometry interacts. As such,
quantum hair is of interest independently of the details as to how
one quantizes gravity.

In concordance with Bekenstein's original suggestion~\cite{Bekenstein},
in many cases it is found that the entropy is simply related to
the area of the event horizon.
\begin{equation}
S = {k A_H\over4\ell_P^2}.
\end{equation}
On the other hand, in many other cases this simple relationship
fails. The pattern, if any, underlying the various results is less
than clear. Consider the following examples.

{$S=(1/4)A$:} The ``entropy = (1/4) area'' law {\em holds} for:
(1) Schwarzschild black holes~\cite{Hawking-Radiation,Gibbons-Hawking},
(2) Reissner-Nordstr\"om black
holes~\cite{Hawking-Radiation,Gibbons-Hawking}, (3) Kerr-Newman
black holes~\cite{Hawking-Radiation,Gibbons-Hawking}, (4) dilatonic
black holes~\cite{Gibbons-Maeda,GHS}, (5) Rotating dilatonic black
holes~\cite{Sen}, (6) generic $(Riemann)^2$ gravity in
$D=4$~\cite{Whitt84}.

{$S\neq(1/4)A$:} The ``entropy = (1/4) area'' law {\em fails} for:
(1) specific examples of $(Riemann)^2$ gravity
($D\neq4$)~\cite{Callan-Myers-Perry,Wiltshire}, (2) generic
$(Riemann)^3$ gravity (D=4)~\cite{Lu-Wise}, (3) specific examples
of  $(Riemann)^4$ gravity~\cite{Myers}, (4) Lovelock gravity
($D\neq4$)~\cite{Myers-Simon,Whitt88}, and (5) various versions of
quantum hair~\cite{DGT,CPW}.

This paper systematizes these results by deriving a general formula
for the entropy in terms of: (1) the area of the event horizon,
(2) the Lorentzian energy density in the classical fields surrounding
the black hole, (3) the Euclideanized Lagrangian describing those
fields, (4) the Hawking temperature, (5) the entropy density
associated with the fluctuations [quantum hair, statistical hair],
and finally (6) the metric. The derivation is particularly transparent,
and the physical interpretation clear, if one temporarily restricts
attention to the spherically symmetric case [zero angular momentum].
In terms of the shape function $b(r)$ and the anomalous redshift
$\phi(r)$ the promised formula reads
\begin{equation}
S = {k A_H\over4\ell_P^2}
  + {1\over T_H} \int_\Sigma \; e^{_\phi} \;
                      \left(\varrho_L-{\cal L}_E\right) d^3r
  + \int_\Sigma {s\over\sqrt{1-(b/r)}} d^3r.
\end{equation}
If no fluctuations are present, ($s=0$, no quantum hair, no
statistical mechanics effects), the issue of the validity of
``entropy = (1/4) area'' law reduces to the question of whether
or not the Lorentzian energy density for the system under consideration
is formally equal to the Euclideanized Lagrangian. As a ``rule
of thumb'':  Lagrangians with quadratic kinetic terms satisfy the
``entropy = (1/4) area'' law. Lagrangians containing $(curvature)^3$
terms and higher typically do not.

The generalization to the case of nonzero angular momentum
(axisymmetric geometry) is straightforward, requiring a little
extra technical machinery in the form of the timelike and azimuthal
Killing vectors, and a suitable invariant integration over the
three-surface defined by taking a constant time slice.

The basic tools to be employed are the relationship between the
thermodynamic functions and the partition function associated with
the ``Wick rotated'' Euclidean section~\cite{Gibbons-Hawking}, and
the Bardeen-Carter-Hawking mass theorem for geometries containing
a timelike Killing vector~\cite{Bardeen-Carter-Hawking}.  The
technical computations are actually relatively simple. Some care
must be taken, however, in carefully navigating through a thicket of
conceptual and definitional issues, and with various subtleties
associated with the shift in signature.

\underbar{Notation:} Adopt units where $c\equiv 1$, but all other
quantities retain their usual dimensionalities, so that in particular
$G\equiv \ell_P/m_P \equiv \hbar/m_P^2 \equiv \ell_P^2/\hbar$. The
metric signature is either $(-,+,+,+)$ or $(+,+,+,+)$ depending on
context. The symbol $T$ will always denote a temperature. The
stress-energy tensor will be denoted by $t^{\mu\nu}$, and its
trace by $t$.

\section{LORENTZIAN TECHNIQUES}
\subsection{The metric, horizon, and Hawking temperature}

In any static spherically symmetric asymptotically flat spacetime
the metric  $g_L$ may without loss of generality be cast into the
form
\begin{equation}
ds^2 = - e^{-2\phi(r)} \left(1 - b(r)/r\right)dt^2
       + {dr^2\over\left( 1 - b(r)/r \right)}
       + r^2(d\theta^2 + \sin^2\theta \; d\varphi^2).
\label{metric}
\end{equation}
The function $b(r)$ will be referred to as the ``shape function'',
while $\phi(r)$ will be referred to as the ``anomalous redshift''
\cite{Visser92}.  Applying boundary conditions at spatial infinity
permits one, without loss of generality, to set $\phi(\infty) = 0$ .
Once this normalization of the asymptotic time coordinate is adopted,
one may interpret $b(\infty)$ in terms of the asymptotic mass
$b(\infty) = 2GM$. This metric has putative horizons at values of
$r$ satisfying $b(r_H) = r_H$. Only the outermost horizon is of
immediate interest.

The Hawking temperature of a black hole is given in terms of its
surface gravity by $k T_H = (\hbar/2\pi) \kappa$.  A brief computation
yields
\cite{Visser92}
\begin{equation}
\kappa = {1\over2 r_H} e^{-\phi(r_H)} \left( 1-b'(r_H) \right).
\end{equation}
This formula receives most of its physical significance after $b'(r_H)$
and $\phi(r_H)$ are related to the distribution of matter by imposing
the Einstein field equations.

The first two Einstein equations are~\cite{Visser92}
\begin{eqnarray}
b' =&& 8\pi G \; \rho \; r^2, \\
\phi' =&& - {8\pi G\over2} {(\rho-\tau)r\over(1-b/r)}.
\label{einstein}
\end{eqnarray}
Instead of imposing the third Einstein equation, observe that (as
is usual) the third equation is redundant with the imposition of
the conservation of stress-energy. Thus one may take the third
equation to be the anisotropic version of the Oppenheimer-Volkoff
equation
\begin{equation}
\tau' = (\rho-\tau)[-\phi' +{1\over2}\{\ln(1-b/r)\}'] - 2(p+\tau)/r.
\end{equation}
Taking $\rho$ and $\tau$ to be primary, one may formally integrate
the Einstein equations:
\begin{eqnarray}
b(r) &=& r_H + 8\pi G \int_{r_H}^r \rho {\tilde r}^2 d{\tilde r}
         = 2GM -  8\pi G \int_r^\infty \rho {\tilde r}^2 d{\tilde r}, \\
\phi(r) &=& {8\pi G\over 2}
            \int_r^\infty {(\rho-\tau)\tilde r\over(1-b/{\tilde r})}
	    d{\tilde r}.
\end{eqnarray}
The transverse pressure $p$ is then determined via the anisotropic
Oppenheimer-Volkoff equation. The Hawking temperature is
\begin{equation}
k T_H = {\hbar\over4\pi r_H} \;
        \exp\left(-{8\pi G\over 2}
	         \int_{r_H}^{\infty}{(\rho-\tau)r\over(1-b/r)} dr
	    \right) \;
        \left( 1 - 8\pi G \;\rho_H \; r_H^2 \right).
\end{equation}
Attempting to determine the entropy by integrating the thermodynamic
relation $dM = T_H dS$ works well in simple cases but in general
quickly leads to an impenetrable morass.  This is about as far as
one can get using Lorentzian techniques. A different method of
attack is called for.

\section{EUCLIDIAN TECHNIQUES}
\subsection{The metric, horizon, and Hawking temperature}

Another way of calculating the Hawking temperature is via the
periodicity of the ``Wick rotated'' Euclidean signature analytic
continuation of the manifold~\cite{Gibbons-Hawking}. Proceed by
making the formal substitution $t\to -it$ to yield a fiducial
Euclidean metric $g_E$:
\begin{equation}
ds^2_E = + e^{-2\phi(r)} \left(1 - b(r)/r\right) dt^2
       + {dr^2\over\left( 1 - {b(r)/r} \right)}
       + r^2(d\theta^2 + \sin^2\theta d\varphi^2).
\end{equation}
In view of the $t$ independence of this metric, this ``Wick rotation''
preserves the mixed components of the Riemann and Ricci tensors:
\begin{eqnarray*}
[Riemann(g_E)]^\alpha {}_\beta {}^\gamma {}_\delta &=&
[Riemann(g_L)]^\alpha{}_\beta {}^\gamma {}_\delta, \\  \relax
[Ricci(g_E)]^\alpha{}_\beta &=&
[Ricci(g_L)]^\alpha{}_\beta, \\
R(g_E) &=& R(g_L).
\end{eqnarray*}
 As is usual,
discard the entire $r<r_H$ region, retaining only the (analytic
continuation of) that region that was outside the outermost horizon
({\it ie:} $r\geq r_H$). A Taylor series expansion about $r=r_H$
shows that the $(r,t)$ plane is a smooth two dimensional manifold
if and only if $t$ is interpreted as an angular variable with period
\begin{equation}
\tau_H = 4\pi r_H \; e^{\phi(r_H)} \;(1-b'(r_H))^{-1} = 2\pi/\kappa.
\end{equation}
Invoking the usual incantations~\cite{Gibbons-Hawking}, this
periodicity in imaginary (Euclidean) time is interpreted as evidence
of a thermal bath of temperature $k T_H = 1/\beta_H = \hbar/\tau_H$,
so that the Hawking temperature is identified as
\begin{equation}
k T_H = {\hbar\over 4\pi r_H} \; e^{-\phi(r_H)} \; (1-b'(r_H)).
\end{equation}
This is the same result as was obtained by direct calculation of
the surface gravity.

\subsection{Helmholtz free energy}

The Helmholtz free energy of an arbitrary statistical mechanical
system is defined in terms of the partition function as
\begin{equation}
F = - k T \; \ln Z.
\end{equation}
For the particular case at hand one writes the partition function as
\cite{Gibbons-Hawking}
\begin{equation}
Z = \int {\cal D}(g,\Phi) \; \exp[-I_E(g,\Phi)/\hbar].
\end{equation}
Here $\Phi$ denotes the generic class of matter fields: fermions,
gauge bosons, Higgs particles, axions, dilatons, etc, etc. The
range of integration runs over all possible matter field configurations,
and over some suitable class of Euclidean metrics.  There is some
confusion as to the class of Euclidean metrics which should be
integrated over in general, but for the present problem it
is sufficient to integrate over all Euclidean metrics $g$ that have
the same topology as the fixed fiducial metric $g_E$, are asymptotically
flat, and are periodic in imaginary time with period $\tau_H =
2\pi/\kappa =\hbar\beta = \hbar/k T_H$~\cite{CPW}.  By adopting
background field techniques one can define an exact decomposition
\begin{equation}
Z = \exp[-I_E(g_E,\Phi_0)/\hbar] \; Z_{\rm fluctuations}.
\end{equation}
Here $g_E$ is the fiducial background metric, $\Phi_0$ denotes the
background matter fields, and $Z_{\rm fluctuations}$ denotes the
contributions to the partition function coming from quantum
fluctuations around the fiducial background --- these fluctuations
can be described by the usual loop expansion.

[Anyone who is worried about the precise class of metrics to
integrate over, or unhappy about invoking background field techniques
can go straight from the definition of the partition function to
the semi-classical limit. Doing so yields an approximation
\begin{equation}
Z \approx \exp[-I_E(g_E,\Phi_0)/\hbar] \; Z_{one-loop}.
\end{equation}
This version of the semiclassical limit handles only one loop
effects in linearized gravitational and matter fluctuations.]

Adopting either of these decompositions one may write
\begin{equation}
F = {k T I_E\over\hbar} + F_{\rm fluctuations}.
\end{equation}
The various contributions to the Euclidean action can be grouped
into three distinct terms
\begin{equation}
I_E(g_E,\Phi_0) =
- {1\over8\pi G} \int_{\partial\Omega} [K] \sqrt{{}_3 g_E} \; d^3x
- {1\over16\pi G} \int_\Omega R \sqrt{g_E} \; d^4x
+ \int_\Omega  {\cal L}_E \sqrt{g_E} \; d^4x.
\end{equation}
These various terms are: (1) the gravitational surface term, to be
integrated over the three surface  at spatial infinity (topology
$S^2\times S^1$), (2) the Einstein-Hilbert term, to be integrated
over the entire Euclidean manifold (topology $S^2\times D^1$), and
(3) the Euclideanized ``matter'' Lagrangian.  Higher order geometrical
terms (e.g. $Riemann^2$), if present, are lumped into the ``matter''
Lagrangian.

The boundary term is easily evaluated:
\begin{equation}
- {1\over8\pi G} \int_{\partial\Omega} [K] \sqrt{{}_3 g_E} \; d^3x
= -{1\over8\pi G} \tau_H (-4\pi G M)
= + {M \tau_H\over2}
= {\hbar\beta M\over2}.
\end{equation}
To evaluate the Einstein-Hilbert term one invokes the Einstein
field equation $G_{\mu\nu} = 8\pi G \; t_{\mu\nu}$. In conformance
with the  conventions already established the Euclidean stress-energy
tensor is defined by setting its mixed components equal to the
mixed components of the Lorentzian stress-energy: $(t_E)^\mu{}_\nu
= (t_L)^\mu{}_\nu$. Consequently, for the trace, $t_E = t_L$. The
subscripts $(E,L)$ will often be omitted if no confusion can arise.
Thus
\begin{eqnarray}
- {1\over16\pi G} \int_\Omega R \sqrt{g_E} \; d^4x
&=& - {1\over16\pi G} \int_\Omega (-8\pi G t) \sqrt{g_E} \; d^4x
= + {1\over2}\int_\Omega t \sqrt{g_E} \; d^4x \nonumber \\
&=& + {1\over2}\int_\Sigma t \; e^{-\phi} 4\pi r^2 dr \;\tau_H
= + {\hbar\beta\over2}\int_\Sigma e^{-\phi}  t \; d^3r.
\end{eqnarray}
Here $\Sigma$ denotes a constant time hypersurface (topology
$S^2\times\Re^+$). Similarly, the matter action can be rewritten as
\begin{equation}
\int_\Omega  {\cal L}_E \sqrt{g_E} \; d^4x
= \hbar\beta\int_\Sigma e^{-\phi} {\cal L}_E \; d^3r.
\end{equation}
Finally, the fact that the Helmholtz free energy is an extensive
quantity justifies the introduction of a free energy density
associated with the fluctuations. This free energy density $f$ is
defined by
\begin{equation}
F_{\rm fluctuations} \; \tau_H = \int_\Omega f \sqrt{g_E} \; d^4x.
\end{equation}
Equivalently
\begin{equation}
F_{\rm fluctuations} = \int_\Sigma  e^{-\phi} f \; d^3r.
\end{equation}
Combining everything
\begin{equation}
F = {M\over2}
  + \int_\Sigma e^{-\phi}
    \left\{ {t\over2} + {\cal L}_E + f  \right\} d^3r.
\end{equation}

\subsection{Bardeen-Carter-Hawking mass theorem}

For a static spacetime the existence of a timelike Killing vector,
together with the use of the Einstein field equations implies
\cite{Bardeen-Carter-Hawking}
\begin{equation}
M = {\kappa A_H\over4\pi G}
  - \int_\Sigma \left\{ 2t_\mu{}^\nu - t \delta_\mu{}^\nu \right\}
              K^\mu d\Sigma_\nu.
\end{equation}
This is a purely geometrodynamic statement in terms of the surface
gravity, the area of the event horizon, and the stress-energy
tensor.  In view of the conventions adopted herein this result
holds equally well in Lorentzian or Euclidean signature. To keep
subsequent formulae more transparent I have reversed the orientation
of the hypersurface $\Sigma$ relative to that adopted by Bardeen,
Carter, and Hawking~\cite{Bardeen-Carter-Hawking}. Thus, with my
conventions, $K^\mu d\Sigma_\mu \mapsto + e^{-\phi} d^3x$ for the
case of spherical symmetry.  Using the relationship between surface
gravity and the Hawking temperature, and using the explicit forms
of the metric and the timelike Killing vector permits this to be
rewritten as
\begin{equation}
M = {k T_H A_H\over2 \ell_P^2}
  + \int_\Sigma e^{-\phi} \left\{ 2\rho + t \right\}  d^3r.
\end{equation}

Resubstituting into the formula for the Helmholtz free energy, in
such a way as to eliminate the integral over the trace of the
stress-energy tensor yields
\begin{equation}
F = M
  - {k T_H A_H\over4 \ell_P^2}
  + \int_\Sigma e^{-\phi} \{ {\cal L}_E + f - \rho \} d^3r.
\end{equation}

\subsection{Thermodynamic Relations}

By definition $F = U - T S$. For an asymptotically flat geometry
the internal energy $U$ is defined to be the asymptotic mass $M$.
Eliminating $F$
\begin{equation}
S = {k A_H\over4 \ell_P^2}
  + {1\over T_H} \int_\Sigma e^{-\phi} \{ \rho - {\cal L}_E - f \} d^3r.
\end{equation}
This is almost the required form. To proceed, note that the $\rho$
occurring above is the {\em total} energy density, and that the way
things have been defined, energy density can arise either from the
classical matter fields surrounding the black hole, or from the
quantum fluctuations, or both. This justifies a split:
\begin{equation}
\rho = \varrho_L + \varrho_f.
\end{equation}
But the energy density in the fluctuations, and the Helmholtz free
energy density in the fluctuations are related by $f = \varrho_f
- T s$, where $s$ is the local entropy density in the fluctuations
and $T$ is the {\em local} temperature. Because the whole system
is at thermal equilibrium at a redshifted temperature $T_H$, the
local temperature varies as
\begin{equation}
T = {T_H\over\sqrt{g_{tt}}} = {T_H \; e^{+\phi} \over\sqrt{1-(b/r)}}.
\end{equation}
Resubstituting everything yields the final result for the entropy
\begin{equation}
S = {k A_H\over4 \ell_P^2}
  + {1\over T_H} \int_\Sigma e^{-\phi} \{ \varrho_L - {\cal L}_E  \} d^3r
  + \int_\Sigma { s \over\sqrt{1-(b/r)}} d^3r.
\end{equation}
This is a very pleasing result which accounts for all known violations
of the ``entropy = (1/4) area'' law in a unified manner. Furthermore,
the result immediately generalizes: instead of considering quantum
fluctuations of the gravitational and matter fields I could just
as easily have dumped a few particles outside the event horizon of
the black hole and proceeded to do ordinary statistical mechanics
in a fixed background geometry. Consequently the fluctuations
discussed in this paper can be thought of as being ordinary
statistical mechanics fluctuations as easily as quantum fluctuations.
The entropy formula derived above applies equally well to dirty
black holes, to classical field configurations, and to stars!
(Subject to the present constraint of spherical symmetry.) Compare
this to the discussion by Gibbons and Hawking~\cite{Gibbons-Hawking}.
Gibbons and Hawking discuss electrovac black holes and perfect
fluid stars.  There is no need in the present formulation for the
effect of the fluctuations, or for the effect of the classical
matter fields, to be constrained to mimic a perfect fluid --- any
generic stress-energy tensor will suffice.

In adding statistical-mechanical hair to the system, one may also
wish to include discussion of the effect of the chemical potential.
There are two compensating modifications.  First note that for the
system as a whole $F = M - T S - \mu_\infty N$. Here $\mu_\infty$ is
the chemical potential as measured at asymptotic infinity, and $N$
is the total number of particles.  Second, for the statistical
mechanical hair $f = \rho_f - T s - \mu n$. Here $\mu$ is the locally
measured chemical potential, and $n$ is the local number density.
Because the whole system is taken to be in chemical equilibrium,
the local chemical potential must be a constant up to a redshift
factor: $\mu = \mu_\infty/\sqrt{g_{tt}}$. The putative additional
contribution to the entropy is proportional to
\begin{equation}
\mu_\infty N - \int_\Sigma e^{-\phi} \;\mu \;n \;d^3r
= \mu_\infty N - \mu_\infty \int_\Sigma n \sqrt{g_3}\;d^3x
= 0
\end{equation}
The formula for the entropy is not disturbed by the addition of a
chemical potential to the system.

Another immediate generalization is that to an arbitrary static,
asymptotically flat, but not spherically symmetric spacetime. The
metric is
\begin{equation}
ds^2 = - e^{-2\Psi} dt^2 + g_{ij} dx^i dx^j
\end{equation}
and the entropy becomes
\begin{equation}
S = {k A_H\over4 \ell_P^2}
  + {1\over T_H} \int_\Sigma e^{-\Psi}
  \{ \varrho_L - {\cal L}_E  \}  \sqrt{g_3} d^3x
  + \int_\Sigma s \sqrt{g_3} d^3x.
\end{equation}
A subtlety is that because I have not placed any energy conditions
on the stress tensor one cannot now invoke the usual proof that
the Hawking temperature is a constant over the horizon. Instead,
constancy of the Hawking temperature over the horizon is now enforced
by the assumption that the system is in thermal equilibrium

A striking feature of the entropy formula is the existence of an
anomalous contribution associated with the interplay between certain
types of classical field and the existence of the heat bath.
Explicitly
\begin{equation}
S_{\rm anomalous}
= {1\over T_H} \int_\Sigma e^{-\phi} \{ \varrho_L - {\cal L}_E  \} d^3r
= {k\over\hbar} \int_\Omega \{ \varrho_L - {\cal L}_E  \} \sqrt{g_E} d^4x.
\end{equation}
In many cases, this anomalous entropy vanishes. In many other cases
it does not.

\section{THE ANOMALOUS ENTROPY}
\subsection{Lagrangians containing only first-order time derivatives}
\subsubsection{Quadratic Kinetic Energy:}

Consider a Lorentzian Lagrangian that is quadratic in first-order
time derivatives. Such a Lagrangian may without loss of generality
be cast in the form
\begin{equation}
{\cal L}_L = {1\over2} g_{ab}(\Phi)\; \dot\Phi^a \dot\Phi^b - V(\Phi).
\end{equation}
The Lorentzian energy density is
\begin{equation}
\varrho_L = \pi_a \dot\Phi^a - {\cal L}_L
          =  {1\over2} g_{ab}(\Phi)\; \dot\Phi^a \dot\Phi^b + V(\Phi).
\end{equation}
On the other hand the Euclideanized Lagrangian is defined by ${\cal
L}_E \equiv - {\cal L}_L(t\mapsto -it)$. For the case under consideration
\begin{equation}
{\cal L}_E
=  {1\over2} g_{ab}(\Phi)\; \dot\Phi^a \dot\Phi^b + V(\Phi)
= \varrho_L.
\end{equation}
Consequently the anomalous  entropy  vanishes, and modulo the
effects of quantum and statistical hair, ``entropy = (1/4) area''.

Examples of this behaviour are the electrovac black holes
(Schwarzschild, Reissner-Nordstr\"om, and
Kerr-Newman~\cite{Hawking-Radiation,Gibbons-Hawking}), as well as
the various variations on the theme of the dilatonic black hole
\cite{Gibbons-Maeda,GHS,Sen}. This observation also applies to the
Lagrangian of the standard model of particle physics, modulo minor
technical fiddles with the Fermi fields.  The recent general
discussion of the ``entropy = (1/4) area'' law by Moss~\cite{Moss}
took the quadratic nature of the kinetic terms as a basic assumption.
Consequently that analysis failed to detect the anomalous $\varrho_L
- {\cal L}_E$ term.

\subsubsection{Generic Kinetic energy:}

Still restricting attention to Lagrangians that are first-order in
time derivatives, suppose the Kinetic energy term to be generic
(subject only to time reversal invariance). Then suppressing field
indices one may write
\begin{equation}
{\cal L}_L = K(\dot\Phi^2,\Phi) - V(\Phi).
\end{equation}
The Lorentzian energy density is
\begin{equation}
\varrho_L
= \pi \dot\Phi - {\cal L}_L
=  K'(\dot\Phi^2,\Phi) \; [2\dot\Phi]\; \dot\Phi
 - K(\dot\Phi^2,\Phi)+ V(\Phi).
\end{equation}
On the other hand, the Euclideanized Lagrangian is
\begin{equation}
{\cal L}_E  =  -K(-\dot\Phi^2,\Phi) + V(\Phi).
\end{equation}
In the difference, $\varrho_L - {\cal L}_E$, the potential energy cancels
\begin{equation}
\varrho_L - {\cal L}_E
=  [2\dot\Phi^2] \; K'(\dot\Phi^2,\Phi)
 - K(\dot\Phi^2,\Phi)
 + K(-\dot\Phi^2,\Phi).
\end{equation}
This looks like a mess. Fortunately, {\em if} the field $\Phi$ is a
physical field, one can use the static nature of the spacetime to
deduce $\dot\Phi=0$.  In this case
\begin{equation}
\varrho_L - {\cal L}_E  =  0 - K(0,\Phi) + K(0,\Phi) = 0,
\end{equation}
and the ``entropy = (1/4) area'' law follows.

\subsection{Lagrangians containing arbitrary order time derivatives}

Independent of the order of time derivatives appearing in the
Lagrangian, the stress-energy tensor may be defined by
\begin{equation}
t^{\mu\nu}(x) =
-{2\over\sqrt{-g}}
{\delta\over\delta g_{\mu\nu}(x)}
   \left[\int_\Omega \sqrt{-g}\; {\cal L}_L \right].
\end{equation}
More explicitly
\begin{equation}
t^{\mu\nu} =
 -2{\delta{\cal L}_L\over\delta g_{\mu\nu}} + g^{\mu\nu} {\cal L}_L.
\end{equation}
Here the symbol $\delta {\cal L}_L/\delta g$ denotes $\partial{\cal
L}_L/\partial g$ plus whatever terms arise from integrating by
parts.  Now $\varrho_L = t^{\hat0\hat0} = t^{tt}/|g^{tt}| = t^{tt}
|g_{tt}|$, so
\begin{equation}
\varrho_L =
 -2g_{tt} {\delta {\cal L}_L\over\delta g_{tt}} - {\cal L}_L
\end{equation}
If one is interested in only physical fields, the static nature of
the spacetime implies, via the vanishing of all time derivatives,
${\cal L}_E \equiv - {\cal L}_L(t\mapsto -it) = -{\cal L}_L$.
Consequently
\begin{equation}
\varrho_L - {\cal L}_E=
 -2g_{tt} {\delta {\cal L}_L\over\delta g_{tt}}.
\end{equation}

The generic breakdown of the ``entropy = (1/4) area'' law in higher
order gravity theories is thus manifest. Typically the variation
with respect to $g_{tt}$ will produce terms such as $R_{t \bullet
t \bullet}$ or such as  $R_{t \bullet \bullet \bullet} R_{t \bullet
\bullet \bullet}$. Without the presence of an accidental zero, the
failure of the ``entropy = (1/4) area'' follows. In agreement with
these observations the law fails for:  $(Riemann)^2$ gravity
($D\neq4$)~\cite{Callan-Myers-Perry,Wiltshire}, $(Riemann)^3$
gravity (D=4) \cite{Lu-Wise}, $(Riemann)^4$ gravity~\cite{Myers},
and Lovelock gravity ($D\neq4$)~\cite{Myers-Simon,Whitt88}.

Accidental zeros of the type alluded to above preserve the ``entropy
= (1/4) area'' law for $(Riemann)^2$ gravity in $D=4$ [${\cal L}
= R + a_1 R^2 + a_2 R_{\mu\nu} R^{\mu\nu} + a_3 R_{\mu\nu\sigma\rho}
R^{\mu\nu\sigma\rho}$]. To see this, note that in four dimensions
the Gauss-Bonnet formula for the Euler characteristic allows one
to rewrite $\int (Riemann)^2$ as a topological invariant plus a
linear combination of $\int (Ricci)^2$ and $\int R^2$. This system
has been analyzed by Whitt~\cite{Whitt84}. The modifications to
the equations of motion are proportional to the Ricci tensor, with
the result that the Schwarzschild solution remains a solution of
the $(Riemann)^2$ system.

\subsection{Topological Lagrangians}

If the Lagrangian contains a topological piece, its contribution
to the anomalous entropy can be calculated trivially. For instance,
in $D=4$ consider the Gauss-Bonnet and Pontrjagin terms
\begin{equation}
{\cal L}_L = {\alpha\over32\pi^2}
              \{ R^{\alpha\beta\gamma\delta} R_{\alpha\beta\gamma\delta}
	         - 2 R^{\alpha\beta} R_{\alpha\beta} + R^2 \}
           + {\beta\over8\pi^2} \{ F^{\mu\nu} \tilde F_{\mu\nu} \}.
\end{equation}
For such topological terms the energy density $\varrho_L$ is zero
by definition. The anomalous entropy reduces to
\begin{equation}
S_{\rm anomalous}
= - {1\over T_H} \int_\Sigma e^{-\phi} {\cal L}_E \; d^3r
= - {k\over\hbar} \int_\Omega {\cal L}_E  \; \sqrt{g_E} d^4x
= - k \{\alpha\chi +\beta p\}.
\end{equation}
This is a simple fixed offset to the entropy generated by the Euler
characteristic and Pontrjagin index of the manifold.  This result
is not exactly surprising and could have been easily deduced from
the original definition of the Helmholtz free energy. If $Z_0$
denotes the partition function excluding topological effects $F =
- k T \; ln Z = k T \{\alpha\chi +\beta p\} - k T \; ln Z_0$.

\section{AXISYMMETRIC SPACETIMES}

The discussion up to the present has, for simplicity, only discussed
the spherically symmetric case. To relax this constraint to merely
require axial symmetry is not particularly difficult. (One needs
to do this in order to be able to  discuss black holes possessing
angular momentum.)

In a stationary axisymmetric asymptotically flat spacetime there
is a unique translational Killing vector $K^\mu$ which is timelike
and normalized to $K^\mu K_\mu = -1$ near spatial infinity. By
abuse of language this is often referred to as the timelike Killing
vector. There is also a unique rotational Killing vector $\tilde
K^\mu$ normalized by demanding that its orbits are closed curves
with parameter length $2\pi$~\cite{Bardeen-Carter-Hawking}.

The fundamental formula for the Helmholtz free energy in terms of the
Euclidean action is recast as
\begin{equation}
F = {M\over2}
  + \int_\Sigma \left\{ {t\over2} + {\cal L}_E + f  \right\}
     \; K^\mu d\Sigma_\mu.
\end{equation}
Here $\Sigma$ is a spacelike hypersurface, tangent to the azimuthal
Killing vector $\tilde K$. The induced 3-metric has volume form
$d\Sigma_\mu$. By construction $\tilde K^\mu d\Sigma_\mu = 0$.

On the other hand, one form of the Bardeen-Carter-Hawking mass
formula now reads~\cite{Bardeen-Carter-Hawking}
\begin{equation}
M = {\kappa A_H\over4\pi G} + 2\Omega_H J_H
  - \int_\Sigma \left\{ 2t_\mu{}^\nu - t \delta_\mu{}^\nu \right\}
              K^\mu d\Sigma_\nu.
\end{equation}
The extra contribution involves the angular momentum of the black
hole $J_H$, and the angular velocity of the event horizon $\Omega_H$.
The angular momentum of the black hole is defined by
\begin{equation}
J_H = +{1\over8\pi G} \int_{\rm horizon} \tilde K^{\mu;\nu} d\Sigma_{\mu\nu}.
\end{equation}
To proceed it is advantageous to further massage the term $\int
t_\mu{}^\nu K^\mu d\Sigma_\nu $. Note that the stress-energy
surrounding the black hole should be rotating ``with'' the black
hole. This notion may be formalized by requiring the stress-energy
tensor to possess a timelike unit eigenvector $V^\mu$, with
corresponding eigenvalue $\rho$. Explicitly
\begin{equation}
t^\mu{}_\nu  V^\nu = - \rho V^\mu.
\end{equation}
This in fact defines the comoving energy density. An observer with
four velocity $V^\mu$ sees no energy flux. By the assumed axial symmetry
the four velocity must be of the form
\begin{equation}
\lambda V^\mu = K^\mu + \omega \tilde K^\mu.
\end{equation}
($\lambda$ is a normalizing factor.) This indicates that, as
expected, the stress-energy surrounding the hole is rotating
``with'' it.  The value of formalizing these notions in this indirect
manner is that one is no longer restricted to the case of a perfect
fluid. ({\em cf}~\cite{Gibbons-Hawking,Bardeen-Carter-Hawking}.)
For the discussion at hand one is interested only in a system in
internal equilibrium. Hence one sets $\omega = \Omega_H$. (Everything
rotates at the same angular velocity throughout the system.)
Repeatedly using the fact that $\tilde K^\mu$ is tangent to the
hypersurface $\Sigma$
\begin{eqnarray}
\int_\Sigma t_\mu{}^\nu K^\mu d\Sigma_\nu
&=& \int_\Sigma t_\mu{}^\nu
    (\lambda V^\mu - \Omega_H \tilde K^\mu )d\Sigma_\nu
    \nonumber\\
&=& \int_\Sigma
    (-\lambda \rho V^\nu - \Omega_H t_\mu{}^\nu \tilde K^\mu) d\Sigma_\nu
    \nonumber\\
&=& -\int_\Sigma \rho( K^\mu + \Omega_H \tilde K^\mu) d\Sigma_\mu
  - \Omega_H \int_\Sigma t_\mu{}^\nu \tilde K^\mu d\Sigma_\nu
    \nonumber\\
&=& -\int_\Sigma \rho K^\mu d\Sigma_\mu - \Omega_H J_{\rm matter}.
\end{eqnarray}
The angular momentum of the matter, $J_{\rm matter}$ is defined in
the usual manner~\cite{Bardeen-Carter-Hawking}
\begin{equation}
J_{\rm matter} = + \int_\Sigma t_\mu{}^\nu \tilde K^\mu d\Sigma_\nu
\end{equation}
For the case of interest (internal equilibrium, $\omega = \Omega_H$)
the Bardeen-Carter-Hawking mass theorem now reads
\begin{equation}
M = {\kappa A_H\over4\pi G} + 2 \Omega_H J_{\rm total}
  + \int_\Sigma  \left\{ 2\rho + t \right\}  K^\mu d\Sigma_\mu.
\end{equation}

As was previously also the case, one can eliminate the integral
over the trace of the stress-energy.  Combining the above
\begin{equation}
F = M
  - {k T_H A_H\over4 \ell_P^2} - \Omega_H J_{\rm total}
  + \int_\Sigma \{ ({\cal L}_E + f)  - \rho \}
                  K^\mu d\Sigma_\mu.
\end{equation}
The relationship between the Helmholtz free energy and the
other thermodynamic quantities is also modified. Including the
effects of angular momentum and a chemical potential $F = M - T S
- \Omega_H J_{\rm total} - \mu_\infty N$.  Here $\Omega_H$ is again
promoted to the status of the angular velocity of the entire heat
bath --- not just the angular velocity of the horizon. Eliminating~$F$
\begin{equation}
S = {k A_H\over4 \ell_P^2} - {\mu_\infty N \over T_H}
  + {1\over T_H} \int_\Sigma \{ \rho -({\cal L}_E + f)\ \}
                  K^\mu d\Sigma_\mu.
\end{equation}
To proceed, repeat the previous trick of splitting the total energy
density into contributions from the fields and from the fluctuations:
$\rho = \varrho_L + \varrho_f$.  The energy density in the
fluctuations, and the Helmholtz free energy density in the
fluctuations, being local quantities, are still related by $f =
\varrho_f - T s -\mu n$. Because the whole system is at thermal
equilibrium, the local temperature and local chemical potential
are redshifted by the normalization parameter $\lambda = || K +
\Omega_H \tilde K||$:
\begin{equation}
T = {T_H \over\lambda}; \qquad  \mu = {\mu_\infty\over\lambda}.
\end{equation}
Then
\begin{eqnarray}
\int_\Sigma \{ \varrho_f - f \}  K^\mu d\Sigma_\mu
&=& \int_\Sigma \{ T s + \mu n \}  K^\mu d\Sigma_\mu
\nonumber\\
&=& \int_\Sigma \{ T_H s + \mu_\infty n \}  (K^\mu/\lambda) d\Sigma_\mu
\nonumber\\
&=& \int_\Sigma \{ T_H s + \mu_\infty n \}  V^\mu d\Sigma_\mu
\nonumber\\
&=&  T_H \int_\Sigma s V^\mu d\Sigma_\mu + \mu_\infty N.
\end{eqnarray}
Resubstituting everything yields the final result for the entropy
\begin{equation}
S = {k A_H\over4 \ell_P^2}
  + {1\over T_H} \int_\Sigma \{ \varrho_L - {\cal L}_E  \} K^\mu d\Sigma_\mu
  + \int_\Sigma s V^\mu d\Sigma_\mu.
\end{equation}
This final result now applies to stationary asymptotically flat
axisymmetric spacetimes. The additional technical machinery required
to go beyond spherical symmetry boils down to the introduction of
appropriate volume forms on the constant time hypersurface $\Sigma$,
together with a suitable definition of the energy density in terms
of a co-rotating observer.

The present version of the analysis also makes it clear that there
is nothing special about $(3+1)$ dimensions.  The entropy formula
continues to hold --- with  suitably defined volume forms --- in
arbitrary dimensionality.

\section{DISCUSSION}

In summary, this paper has exhibited a general formalism for
calculating the entropy of stationary axisymmetric asymptotically
flat dirty black holes. The formalism serves to tie together and
explain in a unified manner a  number of otherwise seemingly
accidental results scattered throughout the literature. The total
entropy can be cleanly separated into contributions from: (1) the
horizon, (2) quantum or statistical hair, and (3) an anomalous
term.

The anomalous entropy is
\begin{equation}
S_{\rm anomalous}
=  {1\over T_H} \int_\Sigma \{ \varrho_L - {\cal L}_E  \} K^\mu d\Sigma_\mu
=  {k\over\hbar} \int_\Omega \{ \varrho_L - {\cal L}_E  \} \sqrt{g_E} d^4x.
\end{equation}
It is certainly a peculiar object, depending as it does on both
the temperature and on the classical background fields
surrounding the black hole.  The vanishing or non-vanishing of
this term correctly retrodicts all known violations and all known
verifications of the naive ``entropy = (1/4) area'' law.

The effects of various types of Lagrangian  can be summarized by
a ``rule of thumb'':  Lagrangians with quadratic kinetic terms
do not contribute to the anomalous entropy. Lagrangians containing
$(curvature)^3$ terms and higher typically do contribute to the
anomalous entropy.

This suggests the following physical picture: Start with the standard
model Lagrangian ${\cal L}_0$.  It does not contribute to the
anomalous entropy. Integration over the quantum fluctuations yields
some quantum hair --- call it $s_0$. Now introduce some energy
scale $\Lambda$ and integrate out the fast modes. This yields some
effective Lagrangian ${\cal L}_{\rm eff}(\Lambda)$. Introducing
this effective Lagrangian into the partition function and integrating
out the remaining slow modes will yield modified quantum hair, call it
$s_{\rm eff}(\Lambda)$. But the effective Lagrangian will contain
$(curvature)^3$ terms and higher  --- and these terms will contribute
to the anomalous entropy.  Now the total entropy should not depend
on where one places the division ($\Lambda$) between fast and slow
modes (after all, it is the same physical theory no matter how one
divides it up).  This suggests that occurrence of anomalous entropy
is to a large extent due to the use of effective Lagrangians and
that moving the division line between fast and slow modes merely
shifts entropy to and fro between the anomalous term and the quantum
fluctuations.  From this point of view all known violations of the
``area = (1/4) entropy'' law can be interpreted as probing the
effect of otherwise uncontrollable high frequency quantum fluctuations
by resorting to the use of some low energy effective Lagrangian.
This physical picture has implications external to the topic of
black hole physics insofar at it indicates the existence of a
general scheme for associating a quantum mechanical entropy with
an effective Lagrangian.  Naturally, if the fundamental theory
contains higher curvature terms, some of the anomalous entropy
should be thought of as intrinsic.

As to the future: I would really like to see an explanation for
this result phrased completely in terms of Lorentzian signature
techniques. The Hawking temperature is already well understood from
a purely Lorentzian point of view, and a similar understanding of
the entropy is clearly desirable.

\underline{Acknowledgements}

This research was supported by the U.S. Department of Energy.  I
wish to thank Fay Dowker for a useful discussion and for bringing
reference \cite{Myers-Simon} to my attention.

\newpage

\end{document}